**Multiple polarization orders in individual twinned colloidal nanocrystals of centrosymmetric HfO$_2$**

Hongchu Du,[1,2,6,*] Christoph Groh,[3] Chun-Lin Jia,[1,4,5] Thorsten Ohlerth,[3] Knut W. Urban,[1,4,6] Rafal E. Dunin-Borkowski,[1,4,6] Ulrich Simon,[3,6] Joachim Mayer[1,2,6]

[1]Ernst Ruska-Centre for Microscopy and Spectroscopy with Electrons, Forschungszentrum Jülich GmbH, 52425 Jülich, Germany

[2]Central Facility for Electron Microscopy, RWTH Aachen University, 52074 Aachen, Germany.

[3]Institute of Inorganic Chemistry, RWTH Aachen University, 52054 Aachen, Germany

[4]Peter Grünberg Institute, Forschungszentrum Jülich, 52425 Jülich, Germany

[5]School of Microelectronics and State Key Laboratory for Mechanical Behaviour of Materials, Xi'an Jiaotong University, 710049 Xi'an, China

[6]Fundamentals of Future Information Technology, Jülich Aachen Research Alliance, 52425 Jülich, Germany

*Correspondence: h.du@fz-juelich.de

**Summary**

Spontaneous polarization is essential for ferroelectric functionality in non-centrosymmetric crystals. High-integration-density ferroelectric devices require the stabilization of ferroelectric polarization in small volumes. Here, atomic-resolution transmission electron microscopy imaging reveals that twinning-induced symmetry breaking in colloidal nanocrystals of centrosymmetric $HfO_2$ leads to the formation of multiple polarization orders, which are associated with sub-nanometer ferroelectric and antiferroelectric phases. The minimum size limit of the ferroelectric phase is found to be ~4 $nm^3$. Density functional theory calculations indicate that transformations between the ferroelectric and antiferroelectric phases can be modulated by lattice strain and are energetically possible in either direction. The results of this work provide a route towards applications of $HfO_2$ nanocrystals in information storage at densities that are more than an order of magnitude higher than the scaling limit defined by the nanocrystal size.

**Introduction**

Spontaneous polarization in ferroelectric materials is achieved by symmetry breaking, either globally in a paraelectric-to-ferroelectric phase transition upon cooling,[1] as for the great majority of ferroelectrics, or locally at boundaries at which spatial periodicity is disrupted.[2] Evidence for polarization orders at grain boundaries,[3] antiphase boundaries[4–6] and ferroelastic domain boundaries[7] has been reported in incipient ferroelectric and antiferroelectric perovskite oxides. However, polarization orders that arise from symmetry breaking at boundaries in centrosymmetric materials other than perovskite oxides have rarely been observed.

Current interest in ferroelectrics results in large part from their applications in emerging non-volatile ferroelectric random-access memories (FeRAMs).[8] FeRAMs that are based on traditional ferroelectrics, such as $PbZr_xTi_{1-x}O_3$, face challenges in the fabrication of smaller functional elements and three-dimensional (3D) architectures for high capacity and scalability,[9,10] limiting present FeRAM technologies to 130 nm and above[10]. Recently, nanocrystalline films of doped and pure $HfO_2$ have been reported to show ferroelectric behavior at thicknesses below 10 nm.[11,12] As the integration of $HfO_2$ as a high permittivity dielectric is well established in Si technology, $HfO_2$-based ferroelectrics offer the promise of a new generation of FeRAMs.[9,10]

$HfO_2$ exhibits a rich variety of thermodynamically stable and metastable phases. It undergoes successive transitions between monoclinic (*M*), tetragonal (*T*) and fluorite-type cubic (*C*) crystal structures at 2000 and 2800 K.[13] In all three structures, $HfO_2$ is centrosymmetric and exhibits no ferroelectric polarization. The observed ferroelectric behavior in $HfO_2$ thin films has been attributed to the formation of a metastable polar orthorhombic ($O_{FE}$) phase that has space group $Pbc2_1$, or $Pca2_1$ in a different setting.[14] It has been reported that this orthorhombic phase evolves as a result of the non-equilibrium conditions that exist in nanocrystalline films, in which mechanical constraints, electric field cycling, and alloying have been found to alter phase behavior in a complicated way.[11,15–18] Such intricate effects in films that comprise nanocrystallites with a rich variety of similar crystallographic structures complicate the understanding and hence optimization of ferroelectric properties in $HfO_2$. These complications are further exacerbated because high-angle annular dark-field scanning transmission electron microscopy (HAADF STEM) was used in most previous atomic resolution studies.[19] As the contrast in HAADF STEM images of $HfO_2$ reveals primarily the more strongly scattering Hf atoms but not the more weakly scattering oxygen atoms, local information about polarization is not readily resolvable in either magnitude or direction.

In contrast to nanocrystalline films, polarization orders in colloidal $HfO_2$ nanocrystals have not been reported in the literature.[20–24] Here, we study individual colloidal nanocrystals of $HfO_2$ using state-of-the-art spherical and chromatic aberration corrected transmission electron microscopy (TEM). By employing the negative spherical aberration imaging (NCSI) technique,[25,26] we image oxygen atoms in addition to Hf atoms directly in $HfO_2$ for the first time. We measure their positions atom-by-atom and quantify the polarization unit-cell by unit-cell in individual nanocrystals. We reveal that symmetry breaking at twin boundaries leads to multiple polarization orders that are associated with the known ferroelectric $O_{FE}$ phase and a novel, previously unreported, antiferroelectric phase within individual nanocrystals.

**Results**

In order to facilitate the understanding of atomic details of $HfO_2$, Figures 1A and 1B shows schematic diagrams of the structures of the inversion-symmetric $M$ and ferroelectric $O_{FE}$ ($Pbc2_1$) phases of HfO$_2$, respectively, projected onto the *ac* plane. *Pbc*2$_1$, instead of the standard *Pca*2$_1$ setting for the same space group, is used throughout this work, so that the directions of the normal to (100), the **b** and the **c** axes of the *M* and *O*$_{FE}$ structures coincide, for simplicity. Both structures contain two non-equivalent oxygen atoms, which have three-fold (O1) and four-fold (O2) coordination, respectively, with Hf atoms. Within a unit cell, projected O1-O2-O1 positions appear in straight lines in the *M* structure (Figure 1A) and as arcs in the *O*$_{FE}$ structure (Figure 1B). In the latter case, the geometric center (blue circle) of the two marked hafnium atoms (blue dotted line) is displaced parallel to the *c*-axis direction by distance $\delta$ relative to that of the eight surrounding oxygen atoms (red circle) (Figure 1B). The resulting electric dipole gives rise to a spontaneous polarization **P**$_s$. In contrast, the two geometric centers (red and blue circles) coincide in the *M* structure (Figure 1A).

Colloidal nanocrystals of *M* HfO$_2$ were synthesized using a solvothermal method.[27] The majority of the nanocrystals were found to be twinned, often multiply, and to be slightly elongated with widths and lengths of approximately 4 and 9 nm, respectively (Figure S1). An atomic-resolution image of a nanocrystal was recorded along the [010] crystallographic direction using the NCSI technique,[25,26] which provides highly localized bright contrast at the positions of atomic columns on a dark background (Figure 1C). A magnified view of the region marked by a white dashed rectangle is shown in Figure 2A. In this image, the Hf atoms (highest intensity), two closest neighboring O2 oxygen atoms (medium intensity), and O1 oxygen atoms (lowest intensity), can all be identified. The structure of this region shows twins, whose boundaries are (200) planes.

The ability to observe all types of atomic columns directly shows that the structure across the twin boundaries (Figure 2A, dashed lines) matches the *O*$_{FE}$ phase (Figure 1B) oriented along or opposite to the [010] direction. Depending on whether the O1-O2-O1 positions follow straight lines or arcs, the structures of twins in the *M* phase (Figure 2A) can be assigned to either the *M* or the $O_{FE}$ phase (Figure 2B) in the labelled blocks. Block 9 marks a single-unit-cell (0.5 nm) thin lamella of the $O_{FE}$ phase extending along the viewing direction (domain), which is confined between two regions of the *M* phase (blocks 6–8 and 10). Three successive 180° ferroelectric domains of the $O_{FE}$ phase are present in blocks 3, 4, and 5. These observations indicate the formation of multiple polarization orders, i.e., polar and antipolar states, corresponding to the known $O_{FE}$ phase and a novel, previously unreported, antiferroelectric phase (discussed below) at twin boundaries in individual centrosymmetric $HfO_2$ nanocrystals.

Our interpretation is based on an iterative two-step optimization procedure, which involves comparing the experimental TEM image with image simulations. Since the lattices in a given block are periodic and free of defects in the vertical direction, the image in Figure 2B was averaged in this direction to reduce noise and the influence of local distortions. The fit between the averaged experimental image (Figure 2C) and a best-fitting image (Figure 2D) simulated using the final optimized imaging parameters (Table S1) is excellent. The specimen thickness determined from the fit is 2.1 nm. The final refined structure obtained from the optimization procedure is chemically reasonable in terms of the O1-Hf and O2-Hf bond lengths (Table S2).

Local polarization was mapped by using a combination of precisely measured atom positions[28] and calculated Born effective charges (Table S3) obtained *via* the Berry phase method[29] based on density functional theory (DFT). The resulting polarization map (Figure 3A) reveals that only regions of the $O_{FE}$ phase are spontaneously polarized. It confirms that block 9 is a single ferroelectric domain with an upward polarization, whereas blocks 3, 4, and 5 are 180° ferroelectric domains with antiparallel polarizations, *i.e.*, they are in an antipolar state. The volume of the ferroelectric domain (block 9) is only ~4 $nm^3$, based on the observed 0.5 nm × 4.0 nm dimensions and the inferred specimen thickness of 2.1 nm. This volume provides a size limit for polarization that is more than an order of magnitude smaller than the volume of the nanocrystal. Based on the polarization map, the averaged value of the vertical ($P_y$) component of $P_s$ is 30–40 µC cm$^{-2}$ in the ferroelectric phase and close to zero in the monoclinic regions (Figure 3B). These values agree with those calculated using atomic positions from the final refined structure *via* image matching. The measured value of $P_s$ is smaller than the theoretical value for the bulk $O_{FE}$ phase calculated on the basis of the relaxed structure obtained either using Born effective charges (68 µC cm$^{-2}$) or directly using the Berry phase approach (73 µC cm$^{-2}$). The horizontal ($P_x$) component is close to zero in all regions. This quantitative polarization analysis confirms the formation of multiple polar and antipolar orders in the nanocrystal (Figure 3).

Figure 4A shows details of the atomic structure of the ferroelectric domain in block 9 in Figure 2D. A perspective view of the corresponding structural model is shown in Figure 4B. The structure of the $O_{FE}$ phase has a two-fold screw axis ($2_1$) along the *c* axis passing through the center of the unit cell, corresponding to a 180° rotation (red arrow, Figure 4B) followed by a translation of half of the lattice parameter *c* around and along this axis. The twofold screw symmetry relates the structure on the left side to that on the right side relative to the (200) plane (marked in blue in Figure 4B) of the $O_{FE}$ lattice, or *vice versa*. Interestingly, this symmetry relationship not only applies to the $O_{FE}$ phase in block 9, but also relates the *M* phase structure in block 8 to that in block 10.

The application of the two-fold screw symmetry operation to the structure on either the left or the right side with respect to the (200) plane of the *M* structure results in the formation of a twin boundary, whose structure is exactly the $O_{FE}$ phase, with either downward or upward polarization (Figure 4C). This behavior further confirms that a two-fold screw twin operation coincides with the two-fold screw axis in the ferroelectric $O_{FE}$ structure, as discussed above. Furthermore, regular repetition of this twin operation (corresponding to polysynthetic twinning) on a unit-cell basis in the $M$ phase results in the formation of consecutive 180° ferroelectric domains in blocks 3, 4, and 5. Thus, twinning *via* a twofold screw operation breaks centrosymmetry and gives rise to multiple polarization orders in individual nanocrystals of the centrosymmetric $M$ phase of $HfO_2$.

The atomic structure of the unit-cell-wide 180º ferroelectric domains (blocks 3, 4, and 5 in Figure 2D) corresponds to a novel, previously unreported, antiferroelectric orthorhombic phase of space group *Pbca* with centrosymmetry, denoted below as $O_{AFE}$ (Figures 5A and 5B). This phase differs from the known high-pressure orthorhombic phase $(O_I)^{30}$ of the same space group, as manifested by the configuration of oxygen and hafnium atoms (Figure S2). The $O_{AFE}$ is present in approximately 13% of the twinned nanocrystals. An $O_{AFE}$-phase-dominant nanocrystal is shown in Figure S3. The unit cell of the novel $O_{AFE}$ phase comprises two unit cells of the $O_{FE}$ phase, with spontaneous polarizations that are antiparallel to each other (Figure 5B). As a result, the spontaneous polarization of the $O_{AFE}$ unit cell is zero overall. These features are key ingredients of a Kittel-type antiferroelectric.[31]

DFT calculations were performed to understand the structural stability. The DFT-calculated total energy of the $O_{AFE}$ structure is 63 and 54 meV per formula unit (f.u.) lower than that of the $O_{FE}$ and $O_I$ phases, respectively (Figure S4 and Tables S4–S7). Unlike the $O_{FE}$ phase, the experimentally-observed lattice parameters of the $O_{AFE}$ phase agree well with those of the DFT-relaxed structure (Table S5). As a result of the chemical similarity between Hf and Zr, a new $O_{AFE}$ phase for ZrO$_2$ can be predicted. The total energy of the $O_{AFE}$ structure of ZrO$_2$ (Figure S4 and Table S8) is 27 meV f.u.$^{-1}$ relative to the corresponding $M$ phase, being almost the same as that of HfO$_2$.

In order to explore the possibilities of structural phase transitions, we calculated the minimum energy transformation path between the $O_{AFE}$ and $O_{FE}$ phases using normal nudged elastic band (NEB) and generalized solid-state NEB (G-SSNEB) methods[32] based on DFT calculations. The DFT-relaxed structure of the $O_{AFE}$ phase has a larger cell volume than that of double cells of the $O_{FE}$ phase. The ratio of the relaxed lattice parameters ($a_{AFE}$ : $2a_{FE}$ / $b_{AFE}$ : $b_{FE}$ / $c_{AFE}$ : $c_{FE}$) is 1.01 / 0.98 / 1.05. For the NEB path with fixed cell parameters of the $O_{AFE}$ phase, the lattices at reaction coordinates towards to the $O_{FE}$ phase undergo tensile strain in the *c*-axis direction, with compressive strain in the *b*-axis direction (Figure 5C). Fixed cell parameters of the $O_{FE}$ phase lead to converse strains. The lattices in the path calculated using the SSNEB method undergo zero strain, as both atomic and unit cell parameters were relaxed. As shown in Figure 5C, by varying the lattice strain the energy barriers can be reduced to 72 and 41 meV f.u.$^{-1}$, which is comparable to that (~80 meV f.u.$^{-1}$) required to reverse the polarization of the $O_{FE}$ phase.[33] The transformation between the $O_{AFE}$ and $O_{FE}$ phases is therefore predicted to be energetically possible in either direction through an orthodromic intermediate phase, and a field-induced antiferroelectric to ferroelectric transition can be expected to manifest antiferroelectric behavior.

**Discussion**

Our observations reveal that a two-fold screw twin operation in the (200) plane, instead of a mirror operation with respect to the (100) plane,[20–22] explains the geometry of twinning observed in bulk and nanocrystalline monoclinic HfO$_2$. In bulk HfO$_2$, this type of twinning is known to result from a martensitic transformation from the tetragonal to the monoclinic phase upon cooling during melt synthesis.[34,35] It has also been observed in colloidal crystals, in which the martensitic transformation was reported to take place during growth following nucleation in the tetragonal phase.[20–22]

A tetragonal-to-monoclinic transformation in a nanocrystal as a whole cannot explain the observed regularly-repeating twinning on a unit-cell basis. In our study, we frequently observed thin layers of the tetragonal phase at the {100} surfaces of twinned nanocrystals (Figure S5). The observed surface tetragonal phase is likely to have formed during the solvothermal synthesis conducted in an autoclave, as an increase in both temperature and pressure favors the formation of the tetragonal phase. This surface tetragonal phase is expected to transform to the monoclinic phase when another layer of the lattice grows on top of it. This inference is supported by the fact that we did not observe the tetragonal phase inside any nanocrystals. Layer-by-layer growth, the formation of a surface tetragonal phase under certain kinetic conditions and a surface tetragonal-to-monoclinic phase transition provide a reasonable explanation for the observed multiple twins in individual monoclinic nanocrystals.

In-plane compressive strains were predicted by atomistic calculations to induce polarization normal to strains in nanodots of conventional ferroelectric perovskite oxides.[36] The $b$ lattice parameter of the $O_{FE}$ phase inferred from the TEM observations is smaller than that of the relaxed structure, which suggests the presence of compressive strain. Conversely, this compressive strain favors the antipolar $O_{AFE}$ phase (Figure 5C, NEB with fixed cell parameters of the $O_{AFE}$ phase), which explains why the neighboring $O_{FE}$ states in the $a$ axis direction are antiparallel to each other, i.e. the $O_{AFE}$ phase. However, compressive stain cannot explain why the $O_{AFE}$ phase is less frequently observed in twinned nanocrystals. This observation can be understood by considering the kinetics of twinning. First, the surface tetragonal phase is formed only in certain circumstances, as indicated by our TEM observations. Second, not all tetragonal to monoclinic transformations lead to twinning. This combination of two kinetic factors makes the probability of consecutive unit-cell twinning low. Heterogeneous phase formation during growth results in sequential multiple twinning events and, consequently, the coherent intergrowth of $M$, $O_{FE}$ and $O_{AFE}$ phases, giving rise to multiple polarization orders. These polarization orders are fundamentally different from those reported in model antiferroelectric PbZrO$_3$, in which uncompensated thermodynamically stable antiferroelectric states result in finite polarization.[6,37] In contrast, novel $O_{AFE}$ and $O_{FE}$ phases that accommodate the polar and antipolar states are metastable phases of HfO$_2$. The observed ferroelectric and antiferroelectric phases at twin boundaries support the long-proposed notion of a twin plane as a structure-building entity.[38]

A clear picture emerges from our results about how multiple polarization orders form in individual nanocrystals of centrosymmetric materials *via* symmetry breaking by twinning. It is particularly relevant for understanding phase evolution in HfO$_2$ nanocrystalline films that comprise a mixture of grains with different phases. This knowledge is important for the design and optimization of FeRAM devices. In order to understand phase evolution, one should not only take the heterogeneity of grains into account, but also the effects of symmetry breaking on a sub-nanometer-scale within individual grains. *In situ* heating experiments in the TEM have indicated that twin boundaries can mediate monoclinic to tetragonal transitions.[22] However, the structure of the twin boundaries was misinterpreted as being associated with mirror planes due to the missed information about the oxygen atoms.[20–22] Our results reveal all the types of atoms of twin boundaries directly, and hence coherent intergrowth of the $M$, $O_{FE}$ and $O_{AFE}$ phases, which evolve from tetragonal to monoclinic transitions. They provide accurate experimental atomic structure details for building valid models in material and device simulations and offer the opportunity to model intricate systems that comprise boundaries between twins, domains, grains, and phases in nanocrystalline $HfO_2$ films,[39] so that phase evolution and device performance can be adjusted in a controlled manner.

**Conclusions**

We have used atomic-resolution TEM to clarify how twinning can lead to symmetry breaking in centrosymmetric $HfO_2$ nanocrystals, and thereby result in ferro- and antiferroelectric phases at twin boundaries, giving rise to multiple polarization orders. We have shown that two-fold screw symmetry underlies both the structural relationship and the coherent intergrowth of the nonpolar monoclinic, ferroelectric and antiferroelectric phases of $HfO_2$. Our observations of multiple polarization orders in individual nanocrystals provide a size limit for the stabilization of polarization, beyond the value that would be determined by the volumes of the nanocrystals themselves, as would be the case for monodomain nanocrystals,[40] suggesting that information storage can be realized at densities that are higher than the scaling limit defined by the nanocrystal size. In view of the high chemical similarity between Zr and Hf, these conclusions are expected to also apply to $ZrO_2$. Our results pave the way to making use of symmetry breaking for engineering polarization orders in centrosymmetric materials other than perovskite oxides, and open a wide range of new opportunities for exploring emerging ferroelectric systems that are formed from nonpolar materials through twin engineering.

**Experimental Procedures**

**Materials.** Monoclinic nanocrystals of $HfO_2$ were synthesized via a microwave-assisted solvothermal process.[27] The precursor hafnium chloride (0.6 mmol) was transferred to a 10 ml microwave vial within an argon-filled glovebox. In quick succession, 0.5 ml dibenzyl ether and 4 ml benzyl alcohol were added to the vial under rigorous stirring. After 5 minutes, a clear solution was retrieved. The microwave vial was sealed in an autoclave and a solvothermal reaction was performed with a CEM Discover microwave apparatus held at 60 °C for 5 minutes with stirring, then maintained at 220°C for 3 h. After synthesis, the mixture of two separated phases was transferred to a plastic centrifugation tube and washed with 5 ml diethyl ether by centrifugation at 4000 rpm for 5 minutes. The supernatant (the upper hydrophobic phase) in the tube was discarded and the lower phase with the nanocrystals was washed twice by dispersion in ethanol and precipitation with diethyl ether (12000 rpm, 15 minutes). Thereafter, the nanocrystals were re-dispersed in chloroform under stirring and 0.2 mmol of dodecanoic acid and 0.15 mmol of oleylamine were added until a colorless and transparent suspension was obtained. The nanocrystals were purified by adding acetonitrile, followed by centrifugation and resuspension in chloroform.

**TEM imaging.** The nanocrystals dispersed in chloroform were deposited on an ultrathin carbon film supported by lacey carbon on a 400-mesh copper grid from Ted Pella company. Transmission electron microscopy images were recorded at 200 kV using a Gatan OneView phosphor-CMOS camera on an FEI Titan 50-300 PICO electron microscope equipped with a Schottky field emission electron gun, a probe $C_S$ corrector, and a CEOS $C_C/C_S$-corrector. Multiple frames of images were aligned using cross-correlation techniques and averaged to increase the signal to noise ratio.[41] The third-order and fifth-order spherical aberrations $C_3$ and $C_5$ were tuned to obtain optimal high-resolution contrast.[42]

**Supplemental Information**

Details of TEM image simulation and DFT calculation, supplementary figures and tables.

**Acknowledgments**

We thank Dr. M. Lentzen for fruitful discussions and for a critical reading of the manuscript. We thank Dr. J. Barthel for measuring the MTF of the Gatan OneView camera. We acknowledge support from the Deutsche Forschungsgemeinschaft (DFG) under Grant SFB 917 Nanoswitches and under the core facilities Grant MA 1280/40-1. We are grateful for support from the Simulation Laboratory electrons & neutrons (SLen).

**Author Contributions**

C.G. synthesized the nanocrystals. H.D. performed the TEM experiments, image simulations and DFT calculations, analyzed the data, and interpreted the results. U.S. supervised the synthesis experiments. C.L.J., R.D.B. and J.M. provided guidance on the TEM investigations. H.D. and K.U. drafted the paper. All authors discussed the results, commented on the manuscript, and gave approval to the final version of the manuscript.

**Declaration of Interests**

The authors declare no competing interests.

**Data and materials availability**

All relevant data that support the findings of this paper are available within this article and its Supplementary Information, or from the corresponding author on reasonable request.

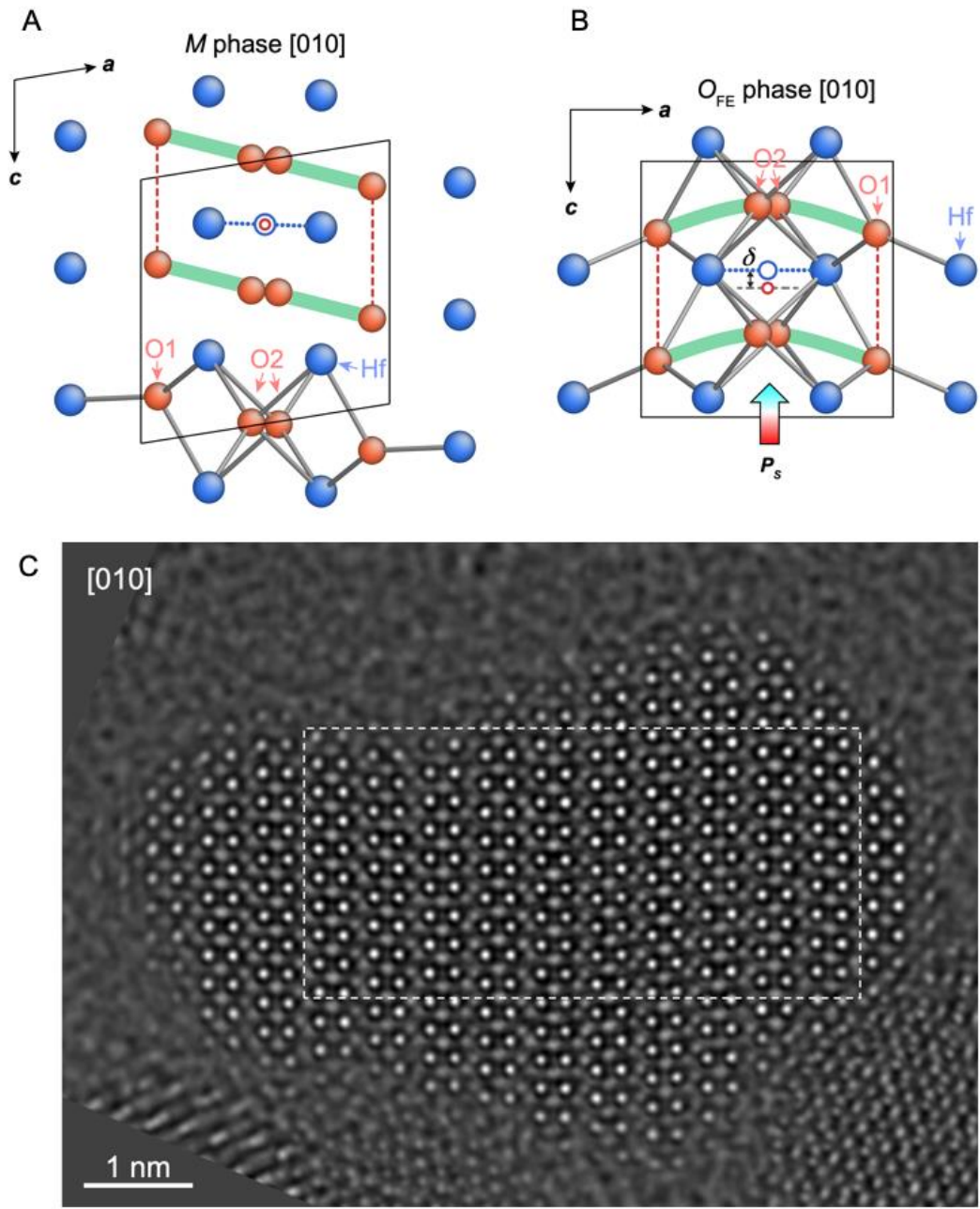

**Figure 1. Schematic structures and TEM image.**
(A) $P2_1/c$ monoclinic (*M*) centrosymmetric phase oriented along the [010] direction. (B) $Pbc2_1$ orthorhombic ($O_{FE}$) ferroelectric phase oriented along the [010] direction. In (A) and (B), black frames indicate the unit cells. O1 and O2 denote oxygen atoms coordinated with 3 and 4 Hf atoms, respectively. The red circle indicates the geometric center of the 8 oxygen atoms on the green lines. The blue circle indicates the geometric center of the two hafnium atoms surrounded by these oxygen atoms. For the $O_{FE}$ structure, a relative displacement δ between the two geometric centers gives rise to a spontaneous polarization ($P_s$). The arrows labelled '$P_s$' indicate the direction of the polarization. (C) Atomic-resolution image of an HfO$_2$ nanocrystal taken along the [010] crystallographic direction using a negative spherical aberration imaging condition.

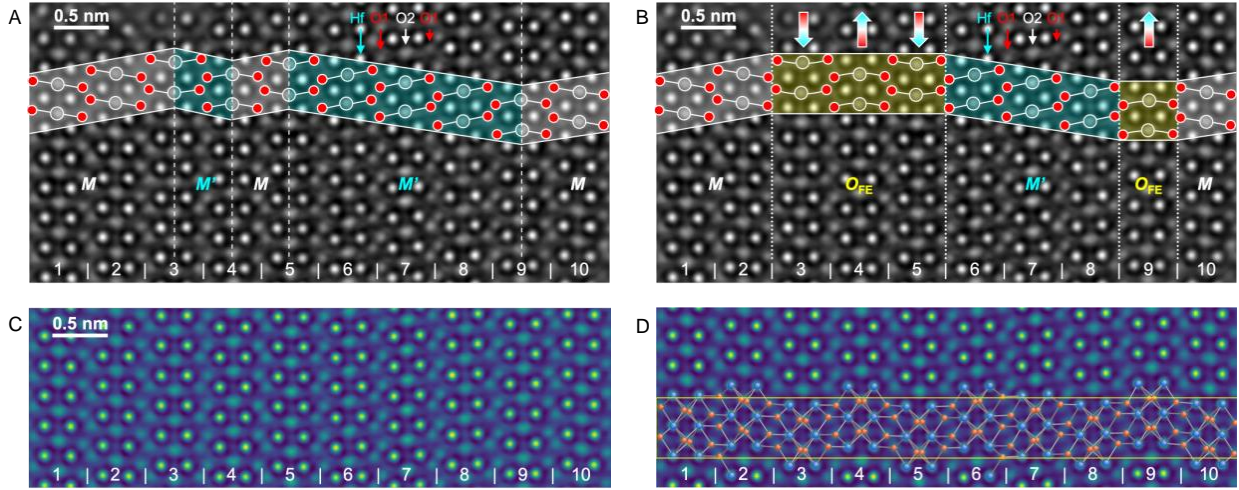

**Figure 2. Multiple twins and polarization orders.**
(A) Magnified view of the region marked by a white dashed rectangle in Figure 1C. The image shows twins of the monoclinic *M* phase with twin boundaries indicated by dashed lines. The structure in an *M* region is related to that in the neighboring *M'* region *via* a twin operation (i.e., a screw axis along the vertical direction in the twin plane). Ten vertical blocks with a horizontal dimension of a single unit cell are labelled 1–10 for ease of reference. (B) The same image shows an intergrowth of the ferroelectric $O_{FE}$ phase (large arrows indicate the direction of the polarization) and the *M* phase. The interfaces between the two phases are outlined by white dotted lines. (C) The image averaged in the vertical direction based on the lattice periodicity. (D) Simulated best-fitting image to the experimental image shown on the same absolute intensity scale as (C). A model of the optimized structure (with Hf in blue, O in red, and O-Hf bonds in grey) is superimposed on the image. The yellow lines indicate the lattice periodicity along the vertical direction.

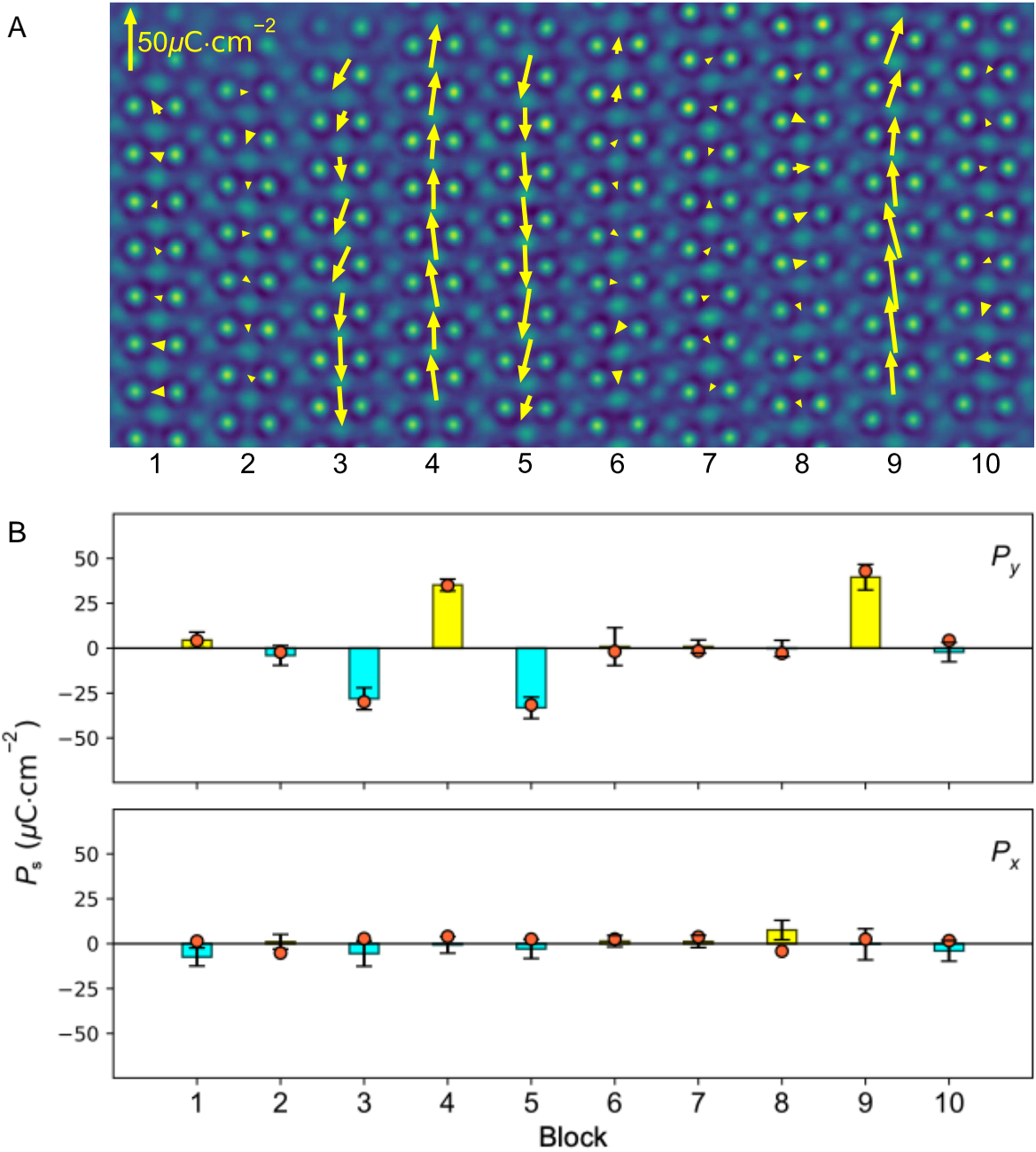

**Figure 3. Spontaneous polarization.**
(A) Map of local spontaneous polarization vectors (arrows) calculated on the basis of half of the unit cell, as illustrated in Figure 1. The length of the arrows represents the modulus of the polarization vectors with respect to the yellow scale bar in the upper left corner. The arrowheads point in the polarization directions. (B) Statistical analysis of vertical ($P_y$, upper panel) and horizontal ($P_x$, lower panel) components of the spontaneous polarization ($P_s$) for structures in each vertical block. The bars represent the averaged values (yellow filled for upward and cyan filled for downward). The error bar represents the standard deviation with respect to the averaged value for each block. The red filled circles denote the values calculated using the atomic positions of the optimized structure corresponding to the simulated image (Figure 2D).

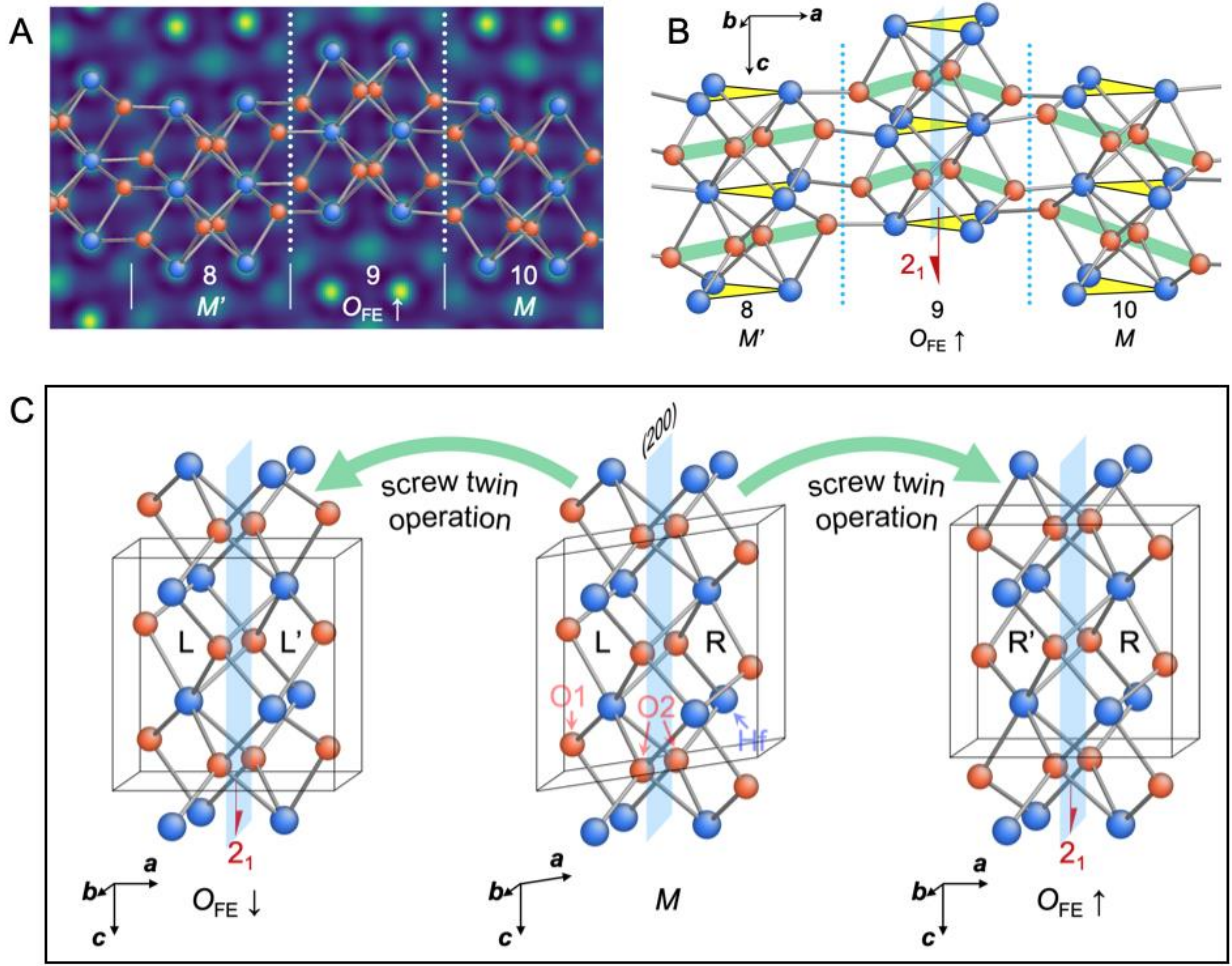

**Figure 4. Twinning to the ferroelectric phase.**
(A) Magnified image of the sub-nanometer ferroelectric domain of the ferroelectric orthorhombic $Pbc2_1$ ($O_{FE}$) phase (block 9, Figure 2D) confined between two parts of the monoclinic $P2_1/c$ ($M$) phase (blocks 8 and10). (B) Perspective view of the structure model for the atomic structure observed in (A). The two-fold screw axis ($2_1$) of the $O_{FE}$ phase is indicated by the red mark. Yellow triangles connecting three Hf atoms facilitate the manifestation of the two-fold screw symmetry. (C) Applying a two-fold screw twin operation ($2_1$) to the structure in either the left part (L) or the right part (R) with respect to the (200) plane of the $M$ phase (middle panel) produces a twin boundary, whose structure is exactly the $O_{FE}$ phase, with downward (left panel) or upward (right panel) polarization. L' and R' indicate the structures after applying the screw twin operation to L and R, respectively. The Hf and O2 atoms coincide almost perfectly for L and R', as well as for R and L'. Arrows close to the '$O_{FE}$' labels mark the directions of polarization.
2323

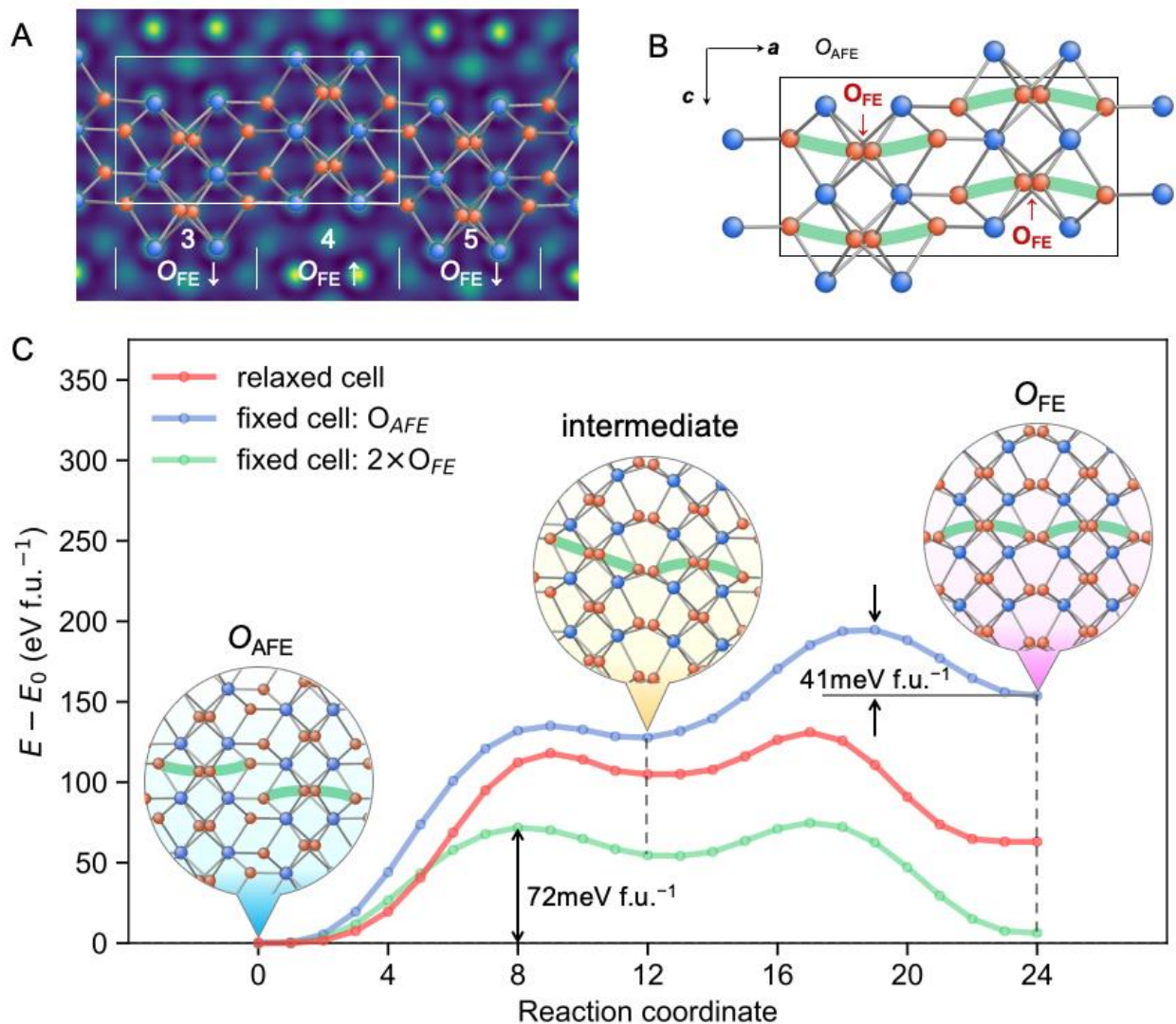

**Figure 5. Novel antiferroelectric phase.**
(A) Magnified view of the 180° domains (lattices in blocks 3, 4, 5 in Figure 2D). (B) Perspective view of the structure model of the novel orthorhombic *Pbca* ($O_{AFE}$) phase, comprising two sets of the ferroelectric orthorhombic *Pbc*2$_1$ ($O_{FE}$) lattices with antiparallel polarization from the observation indicated by the white square in (A). (C) Minimum energy transformation path between $O_{AFE}$ and $O_{FE}$ phases at zero pressure with fixed and free unit-cell calculated using the nudged elastic band (NEB) and solid-state NEB (SSNEB) methods, with climbing images based on density functional theory calculations. The insets show diagrams of the structures of the $O_{AFE}$, $O_{FE}$ and orthodromic intermediate phases along the transformation path.

Supplemental Information

# Multiple polarization orders in individual twinned colloidal nanocrystals of centrosymmetric HfO$_2$

Hongchu Du,[1,2,6,*] Christoph Groh,[3] Chun-Lin Jia,[1,4,5] Thorsten Ohlerth,[3] Knut W. Urban,[1,4,6] Rafal E. Dunin-Borkowski,[1,4,6] Ulrich Simon,[3,6] Joachim Mayer[1,2,6]

[1] Ernst Ruska-Centre for Microscopy and Spectroscopy with Electrons, Forschungszentrum Jülich GmbH, 52425 Jülich, Germany

[2] Central Facility for Electron Microscopy, RWTH Aachen University, 52074 Aachen, Germany.

[3] Institute of Inorganic Chemistry, RWTH Aachen University, 52054 Aachen, Germany

[4] Peter Grünberg Institute, Forschungszentrum Jülich, 52425 Jülich, Germany

[5] School of Microelectronics and State Key Laboratory for Mechanical Behaviour of Materials, Xi'an Jiaotong University, 710049 Xi'an, China

[6] Fundamentals of Future Information Technology, Jülich Aachen Research Alliance, 52425 Jülich, Germany

[*] Correspondence: h.du@fz-juelich.de

## Table of Contents

# 1. Image simulation

Quantum-mechanical and electron-optical image simulations were conducted using Dr. Probe command-line tools[1] and a python interface (https://github.com/FWin22/drprobe_interface). The initial structure model for the image simulations was built based on the projected positions of each type of atom determined by fitting the intensity distribution of the atom contrast using 2D Gaussian peaks[2]. Due to the small separation of the two closest neighboring O2 columns, they appear as a single peak in a TEM image and therefore the fitting position was taken as the average of the positions of the two O2 columns. An iterative two-step optimization procedure was performed by comparing the experimental image pixel-by-pixel with simulated images on an absolute intensity scale. In each iteration of the optimization, all of the instrumental parameters, along with specimen tilt and thickness, were refined first and followed by a second step to optimize the atom positions. The modulation-transfer function (MTF) of the camera was measured using the knife-edge method[3] and was included in the image simulations. We defined the following cost function

$$\sqrt{\frac{\sum_{i=0}^{n}(I_{sim,i}-I_{exp,i})^2/I_{exp,i}^2}{n}} + (1 - \max{(I_{sim} \otimes I_{exp})}),$$

where i is the index of the pixels, n is the total number of pixels, $I_{sim,i}$ and $I_{exp,i}$ are the intensity at the i-th pixel for the simulated and experimental images, respectively, $I_{sim}$ and $I_{exp}$ are the intensities of the simulated and experimental images, respectively, and $\otimes$ is a cross-correlation operation. The first term in the cost function is the weighted root mean square error (wRMSE), while minimization of the second term is equivalent to maximization of the cross-correlation between the experimental and simulated images. The use of wRMSE instead of unweighted RMSE allows pixels of different intensity to be taken into account equally. The imaging parameters, i.e., up to third order aberration coefficients in addition to $C_5$, beam convergence, image spread, objective lens focus spread, specimen tilt and thickness, as well as the lateral atomic positions, were optimized by minimizing the value of the cost function. The atomic positions were optimized with only the first term of the cost function in the second step of each iteration. The final refined parameters are given in Table S1. The approximately zero value of wRMSE (0.032) and the nearly unit maximum value of the cross-correlation (0.99) indicate an almost perfect match between simulation and experiment. Analyses of the bond lengths of O1-Hf and O2-Hf (Table S2) indicate that the final refined structure obtain from the optimization procedure appears to be chemically reasonable in terms of bond lengths.

## 2. DFT calculation

DFT calculations. Structure optimization and total energy calculations were performed by using density functional theory (DFT) within the generalized gradient approximation (GGA) scheme using the Quantum ESPRESSO (QE) software package and ultrasoft pseudopotentials[4,5]. A kinetic energy cut-off of 55 Rydberg was applied to the wavefunctions. The Brillouin zone was sampled with an 8×8×8 Monkhorst-Pack grid for the M and $O_{FE}$ phases, and 4×8×8 for the $O_{AFE}$ phase. Born effective charges $Z^*_{k,ij}$ were calculated by the finite difference of polarization ($\Delta P$) for atomic displacements ($\delta_{k,j}$, 0.01 Bohr) via the Berry phase method[6], with a 12×12×12 Monkhorst-Pack grid using relaxed atomic positions. For the observed structures, the spontaneous polarization was calculated via $\boldsymbol{P}_s = \frac{e}{\Omega}\sum_i \sum_{k,j} Z^*_{k,ij} \delta_{k,j}$, where e is the electron charge, $\Omega$ is the volume of the lattice of interest, i and j are Cartesian directions, $\delta_{k,j}$ is the displacement of atom k in direction j, $Z^*_{k,ij}$ is the Born effective charge tensor[7]. The local value of $P_s$ was calculated using the $Z^*_{k,ij}$ tensors according to the type of atom with respect to the local phase (Table S3). For the $M$ phase, the values of the non-diagonal elements of the $Z^*_{k,ij}$ tensors are non-zero but very small[8], and therefore, only diagonal elements were taken into account for simplicity. The values of $\delta_{k,j}$ were obtained from the atomic column positions determined by Gaussian fitting[2] of the maxima of the experimental TEM image contrast and from the refined structure model obtained through the iterative two-step optimization procedure for matching simulation and experiment. The theoretical values of $P_s$ for the bulk phase were calculated based on the relaxed structure both using Born effective charges and directly using the Berry phase approach, for which the difference is marginal (~7%).

Nudged elastic band (NEB) and general solid-state NEB (G-SSNEB) calculations with climbing images were performed using the TSASE package[9], in combination with GPU-accelerated QE[10] in the Atomic Simulation Environment (ASE)[11] for fixed and free cell minimum energy transformation paths. The initial, final, and an intermediate structure found from the initial relaxed path were fully relaxed (in terms of volume, cell shape and atomic coordinates). The path was split into two paths (initial to intermediate and intermediate to final), so that each NEB/SSNEB calculation would only have a single local maximum. Each split path corresponded to a total of 13 images, including the endpoints. The transformation paths were relaxed until the forces on all images were below 5 meV/Å. The spring constant used in the NEB/G-SSNEB methods was 5.0 eV/Å.

## 3. Supplementary Figure S1–S6

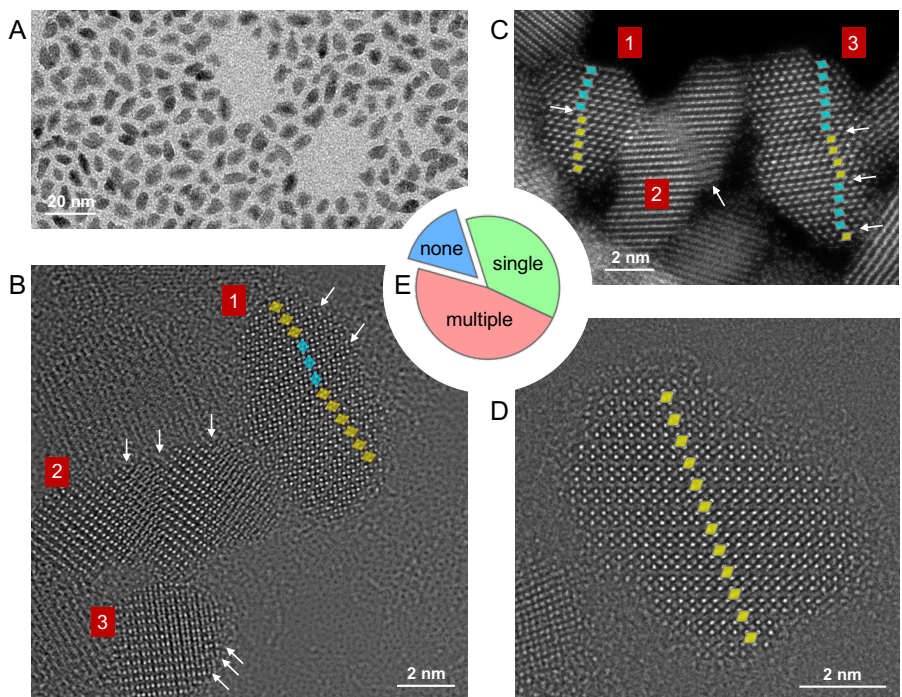

**Figure S1. Transmission electron microscopy (TEM) images of monoclinic HfO$_2$ nanocrystals.**
(A) Overview bright-field diffraction contrast image of slightly elongated HfO$_2$ nanocrystals with widths and lengths of approximately 4 and 9 nm, respectively. The majority of the nanocrystals are twinned, often multiply, on (200) planes. Since Bragg contrast depends sensitively on crystal lattice orientation, a change from bright to dark contrast in individual nanocrystals is often indicative of the presence of twin domains. (B) High-resolution negative spherical aberration TEM image and (C) high-angle annular dark-field (HAADF) scanning TEM (STEM) image of twinned nanocrystals. The HAADF-STEM image was recorded at 200 kV on the PICO electron microscope using a probe semi-convergence angle of 25 mrad and a detector inner collection semi-angle of over 60 mrad. In (B), the nanocrystal labelled '1' is in the [010] orientation, at which O and Hf atoms, twin boundaries and their structures are resolved. In (C), the nanocrystals labelled '1' and '2' are oriented along a direction close to [011]. In this image, the bright dots represent only Hf atoms, as a result of the atomic number (Z) dependence (close to Z$^2$) of HAADF contrast[12]. Nanocrystals '2' and '3' in (B) and '2' in (C) are in other orientations, at which twin boundaries are also visible, but their atomic structures are difficult to interpret. The presence of untwinned nanocrystals is confirmed by the observation of nanocrystals without twin boundaries in [010] and [011] orientations. (D) An example of an untwinned crystals viewed in the [010] orientation. In (B and C), white arrows mark the positions of twin boundaries, which are also identified with the aid of polygons. Based on such images, the probability of the occurrence of nanocrystals with zero, one and multiple twin boundaries is measured to be 16 ± 12 %, 37 ± 15 % and 47 ± 16 % with 95% binomial confidence intervals, respectively, as shown schematically in (E). Only nanocrystals with orientations close to <010> and <011> directions were counted, in order to avoid any ambiguity (See Figure S6).

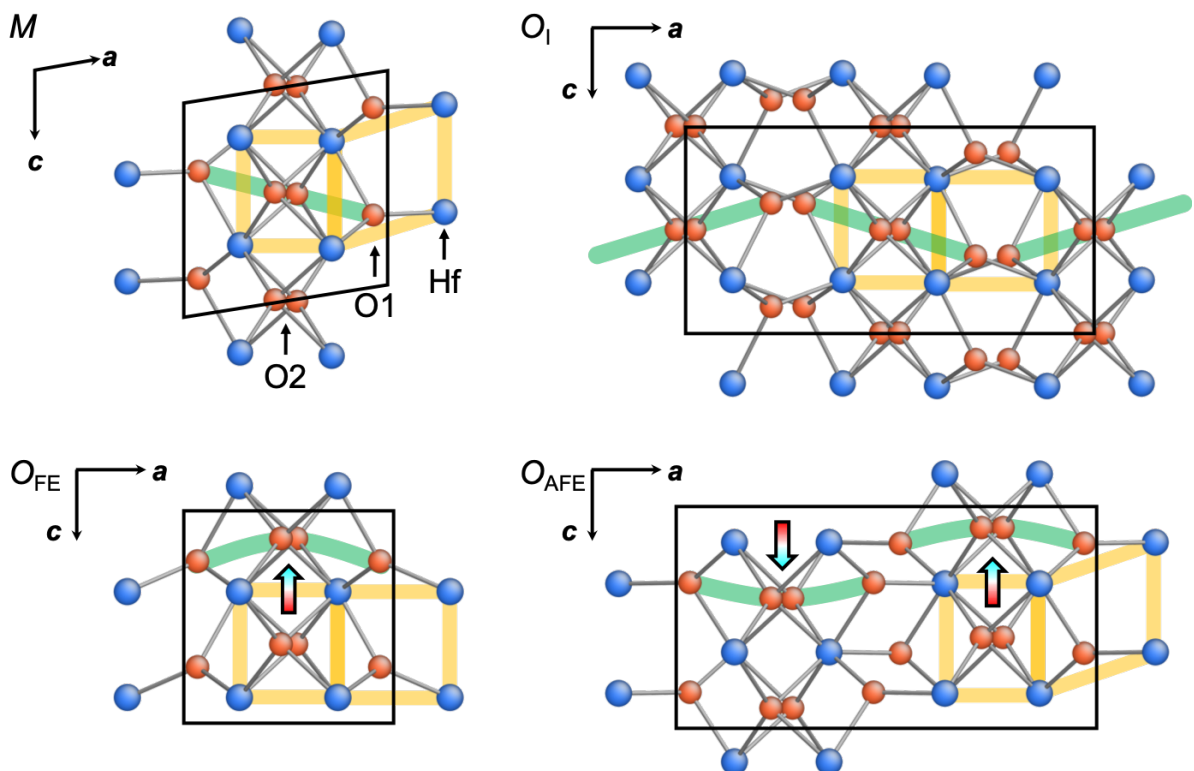

**Figure S2. Comparison of the structures of phases of $HfO_2$.**
$M$: centrosymmetric monoclinic phase with space group No. 14 $P2_1/c$, $O_{FE}$: non-centrosymmetric (polar) orthorhombic phase with space group No. 29 $Pbc2_1$, $O_I$: the reported high-pressure centrosymmetric orthorhombic phase with space group No. 61 $Pbca$ (ICSD 79913). $O_{AFE}$: centrosymmetric orthorhombic phase with space group No. 61 $Pbca$ discovered in this work. The big arrows in the $O_{FE}$ and $O_{AFE}$ structures indicate the polarization directions of the $O_{FE}$ lattice and the antiparallel $O_{FE}$ sublattices, respectively. The green marks indicate the configuration of the projected positions of the O1-O2-O1 appearing in either a straight line or an arc, whereas the yellow marks indicate the configuration of the projected positions of the Hf atoms. With the aid of these marks, a striking similarity of the oxygen lattice between the $M$ and $O_I$ and between the $O_{FE}$ and $O_{AFE}$ structures can be observed, while for the Hf lattice the similarity exists between the $M$ and $O_{AFE}$, and between the $O_{FE}$ and $O_I$ structures.

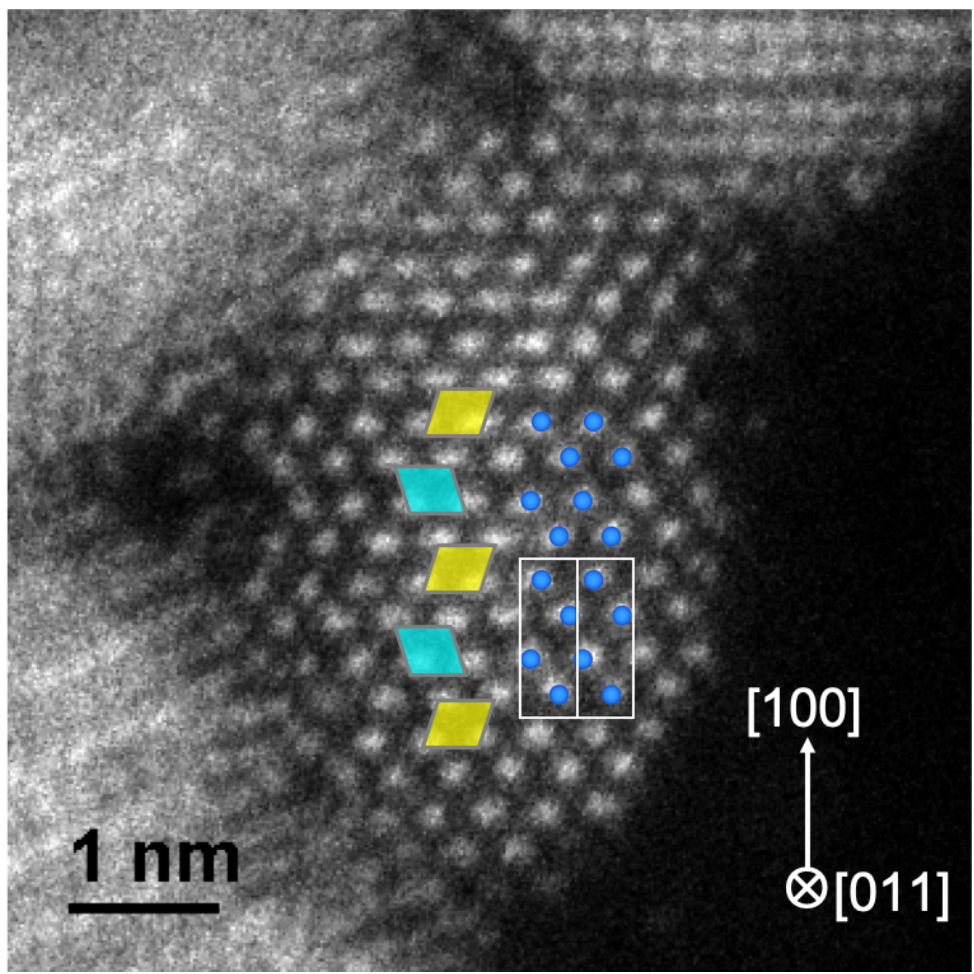

**Figure S3. HAADF STEM image of an $O_{AFE}$ antiferroelectric-phase-dominant HfO$_2$ nanocrystal.**
Polygons of different color indicate different arrangements of projected Hf atoms (blue symbols). White lines indicate the unit cell of the $O_{AFE}$ structure. Oxygen atoms are not visible in the HAADF STEM image and are not indicated for clarity. This nanocrystal comprises a domain of the $O_{AFE}$ phase with a dimension of 2.5 unit cells along the [100] direction. The HAADF STEM image was recorded at 200 kV on the PICO electron microscope using a probe semi-convergence angle of 25 mrad and a detector inner collection semi-angle of over 60 mrad.

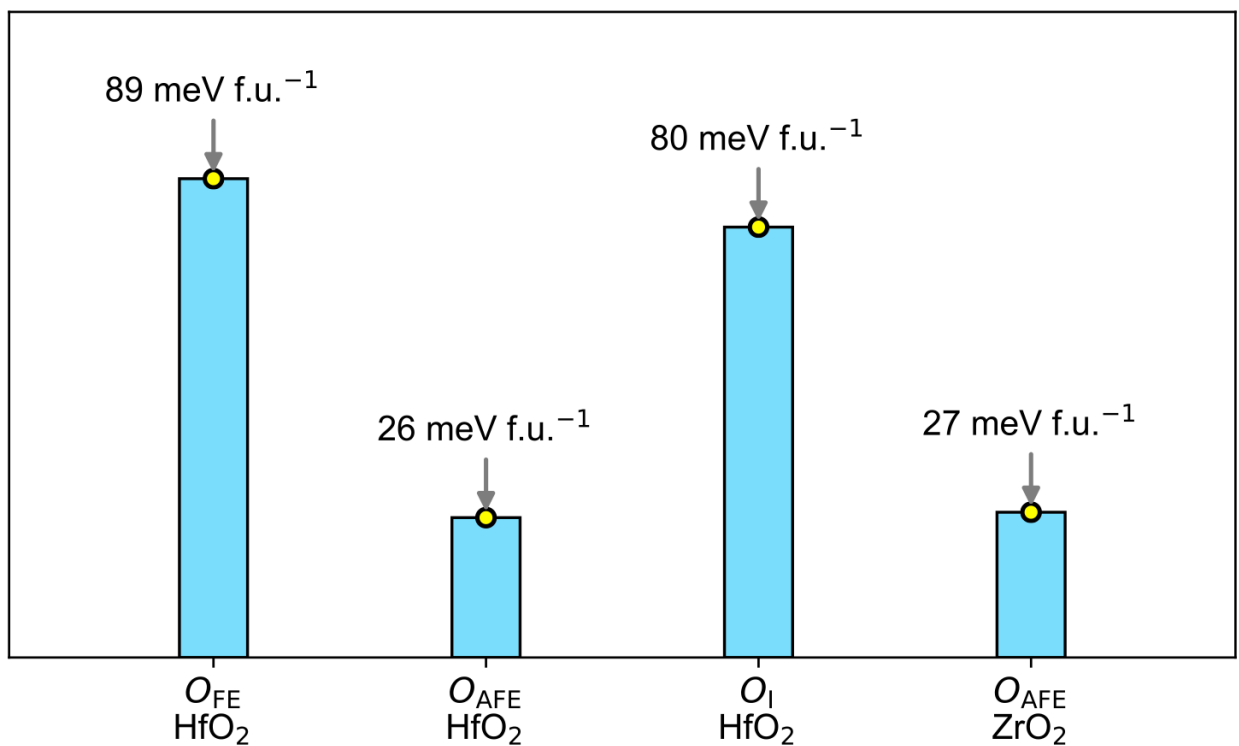

**Figure S4. Total energies of phases of HfO$_2$ and ZrO$_2$.**
The total energies were obtained from DFT calculations based on relaxed structures (Tables S4–S8), relative to that of the monoclinic $P2_1/c$ (*M*) phase.

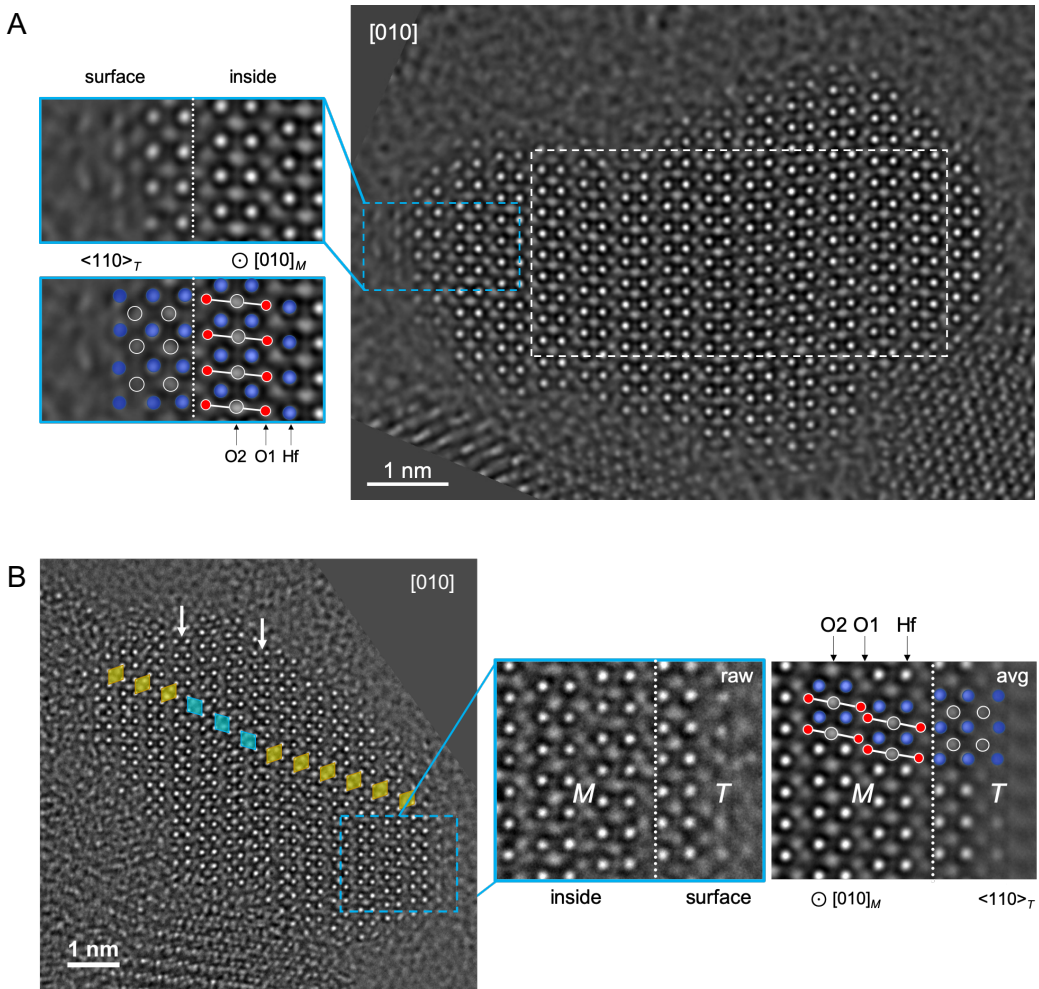

**Figure S5. Surface tetragonal phase of HfO$_2$ nanocrystals.**
Note that all oxygen atoms are four-fold-coordinated in the *T* phase of HfO$_2$. The interface between the surface *T* phase and the internal *M* phase is indicated by white dotted lines. (A) The image on the right is shown in Figure 1C in the main text. The atomic structure of the ($\bar{1}$00) surface, which is marked by a blue dashed rectangle, is magnified and shown in the upper frame on the left side, and reveals a thin layer of the tetragonal-like phase on the surface. The same image is shown in the lower frame on the left side, with atomic columns of Hf, three-fold-coordinated oxygen O1, and four-fold-coordinated oxygen O2, indicated. (B) The image on the left side shows the nanocrystal labelled '1' in Figure S1B, with two twin boundaries indicated by white arrows. The atomic structure of the (100) surface, which is marked by a dashed rectangle, is magnified and shown in the middle frame, revealing a thin layer of the tetragonal-like (*T*) phase on the surface. The image was averaged in the vertical direction based on the periodicity of the lattice. The resulting averaged image is shown on the right side, with atomic columns of Hf, three-fold-coordinated oxygen O1, and four-fold-coordinated oxygen O2, indicated.

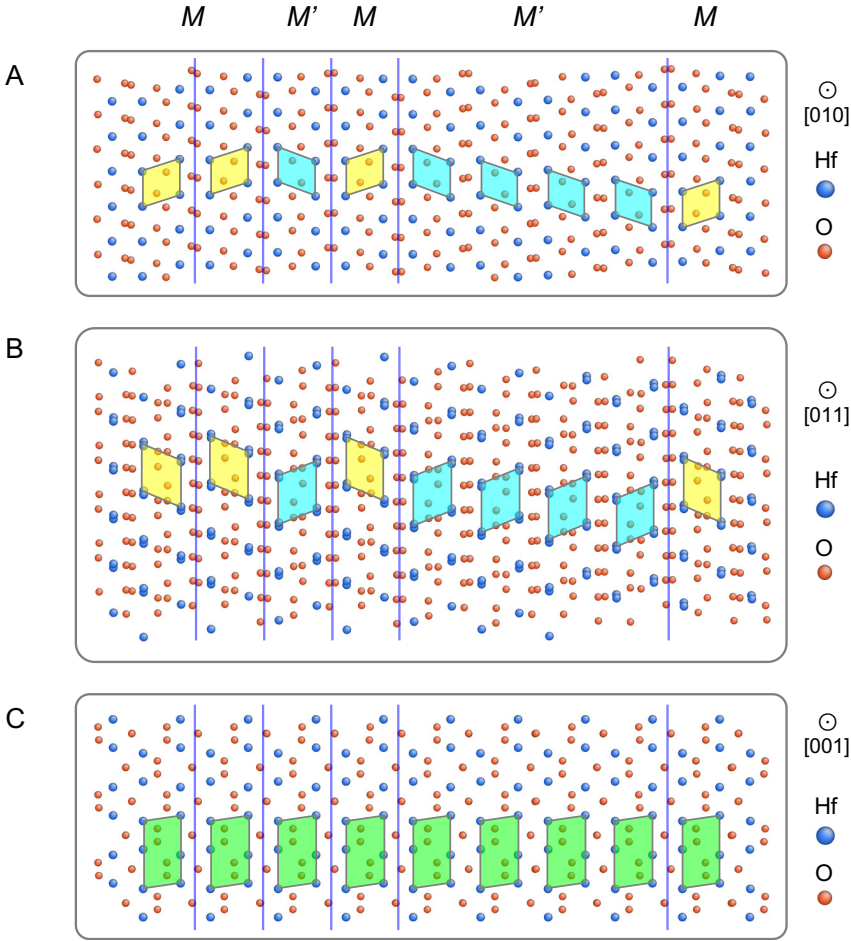

**Figure S6. Orientation dependence of the visibility of twin boundaries.**
(A–C) Atomic arrangements based on the structure model shown in Figure 2d in the main text viewed along (A) [010], (B) [011] and (C) [001] orientations. The structure has four twin boundaries on (200) planes, which are indicated by blue lines. The symbol *M'* indicates each domain that is twinned with respect to a neighboring *M* domain. Bonds are omitted for clarity. The polygons are used as a further aid to identify the positions of twin boundaries in each orientation. The twin boundaries can be imaged unambiguously when nanocrystals are viewed along the <010> and <011> directions, i.e., all directions that are equivalent to [010] and [011], respectively. The projected atomic columns have different arrangements that can be distinguished from one another by using polygon tiling, as shown in (A) for [010] and (B) for [011]. The [010] orientation has the advantage that it allows not only twin boundaries but also ferroelectric orders emerging from symmetry breaking at twin boundaries to be imaged (see Figures 2A and 2B in the main text). When the structure is oriented at or opposite to [001] direction, as shown schematically in (C), the twin boundaries are invisible. Although the twin boundaries are still edge-on to the projection direction, the *M* and *M'* twin domains show the same projected atomic structure. As a result, the twinned structure appears to be perfect when viewed along [001].

## 4. Supplementary Table S1–S8

**Table S1. Final optimized parameters for image simulations.**

| | | |
|---|---|---|
| Acceleration high-voltage | 200kV | |
| Carbon absorption[a] | 0.032 | |
| Specimen thickness | 2.1 nm | |
| Defocus ($f$) | 4.5 nm | |
| Specimen Tilt (x, y) | −0.25 mrad | −0.62 mrad |
| Twofold astigmatism (x, y) | 0.32 nm | 0.12 nm |
| Threefold astigmatism (x, y) | 14.4 nm | −31.7 nm |
| Coma (x, y) | −73 nm | −12.5 nm |
| Third-order spherical aberration ($C_3$) | −8.7 nm | |
| Fourfold astigmatism (x, y) | −1.1142 μm | 0.939 μm |
| Star aberration (x, y) | 0.987 μm | 1.515 μm |
| Fifth-order spherical aberration ($C_5$) | 3.6 mm | |
| Beam convergence half width | 0.45 mrad | |
| Image spread | 16 pm | |
| Focus spread $\Delta f$ [b] | 1.1 nm | |
| Debye-Waller factor of Hf, $B_{Hf}$ | 0.0052 (0.0045[c]) nm² | |
| Debye-Waller factor of O, $B_O$ | 0.0008 (0.0045[c]) nm² | |

[a]There is no vacuum area in the field of view of the recorded image. We therefore used an area of thin carbon instead of vacuum as the reference. We estimated the carbon absorption parameter using an optimization procedure; [b]Instabilities of the specimen on the ultrathin carbon support during the exposure, along with image averaging from multiple frames and positions, may account for the fitted value of 1.1 nm for the focus spread, the standard deviation of the Gaussian focal spread distribution $\Delta f$. With the chromatic aberration $C_C$ for the objective lens (obj.) corrected, the value of $\Delta f$ resulting from instabilities in the high voltage (accelerator) supply $\Delta V$, and the intrinsic energy spread in the electron gun $\Delta E$, are expected to be marginal according to the equation

$$\Delta f = C_c \sqrt{\left(\frac{\Delta V}{V_{acc}}\right)^2 + \left(\frac{\Delta E}{V_{acc}}\right)^2},$$

where $C_C = C_{C,obj.} + C_{C,corrector}$ is close to zero. The focus spread caused by instabilities of the current of the objective lens $\Delta I$ is proportional to $C_{c,obj.} \frac{\Delta I}{I}$, which cannot be compensated by a $C_C$ corrector. In practice, e.g. for the PICO electron microscope, the influence of $\frac{\Delta I}{I}$ is reduced through the improvement of the objective lens power supply, which in turn reduces the focus spread caused by the term $C_{c,obj.} \frac{\Delta I}{I}$; [c]The values in brackets correspond to the Debye-Waller factors of Hf and O from ICSD27313 for the $M$ phase of HfO₂.

**Table S2. Bond lengths between oxygen and hafnium atoms.**

|  | $M$ phase (ICSD 27313) | $O_{FE}$ phase (relaxed structure, table S4) | Refined structure[a] |
|---|---|---|---|
| O1-Hf (nm) | 0.208 | 0.209 | 0.211 |
| O2-Hf (nm) | 0.226 | 0.218 | 0.220 |

[a]Refined structure model shown in Figure 2d in the main text obtained through an iterative two-step optimization procedure for matching simulations to experimental measurements. For the final refined structure obtained using this optimization procedure, the averaged bond length between three-fold-coordinated oxygen (O1) and hafnium atoms (O1-Hf) is 0.211 nm, while it is 0.220 nm between four-fold-coordinated oxygen (O2) and hafnium atoms (O2-Hf). Both values are in line with those of the reference $M$ phase, as well as with those of the DFT-relaxed $O_{FE}$ phase. The agreement between these values suggests that the refined structure obtained from the optimization procedure is chemically reasonable in terms of bond length.

**Table S3. Born effective charge calculated from the relaxed structure using DFT *via* the Berry phase method.**

|    | Monoclinic (*M*), space group: No. 14, $P2_1/c$ [a] | | | Polar Orthorhombic ($O_{FE}$), space group: No. 29, $Pbc2_1$ | | |
|----|------|-------|-------|------|-------|-------|
|    | Hf   | O1    | O2    | Hf   | O1    | O2    |
| xx | 5.31 | −2.95 | −2.36 | 5.41 | −2.91 | −2.50 |
| zz | 4.85 | −2.31 | −2.54 | 4.75 | −2.36 | −2.39 |

[a]The *x* direction is the normal to the (100) plane, while the *z* direction is along the *c* axis.

**Table S4. Lattice parameters for the $O_{FE}$ HfO$_2$ structure.**

| Space group | No. 29, $Pbc2_1$ | | |
|---|---|---|---|
| | Relaxed structure Present work | Ref. | HRTEM measurement |
| a | 5.01183 Å | 5.010 Å | 5.05 Å |
| b | 5.23561 Å | 5.234 Å | 5.17 Å [a)] |
| c | 5.04536 Å | 5.043 Å | 5.29 Å |
| alpha | 90º | | |
| beta | 90º | | 90º |
| gamma | 90º | | |
| Volume | 132.390 Å$^3$ | | 138 Å$^3$ [a)] |
| Atomic fractional coordinate | | Wyckoff | |
| Hf | 0.26618, 0.03316, 0.40000 | 4a | |
| O1 | 0.06638, 0.36474, 0.25556 | 4a | |
| O2 | 0.46352, 0.77158, 0.64852 | 4a | |
| Total energy[b)] | 89 meV per formula unit (f.u.) | | |

[a)]The $b$ axis is not visible in the electron beam direction, and thus the experimental value of the $b$ lattice parameter of the $M$ structure from ICSD 27313 was assumed. Owning to the coherent intergrowth of the $O_{FE}$ and $M$ phases, the value of the $b$ lattice parameter of the $O_{FE}$ structure is the same as for the $M$ structure. Hence, the unit cell volumes of the two structures are the same (Table S6); [b)]Relative to the monoclinic $P2_1/c$ ($M$) phase.

**Table S5.** Lattice parameters for the $O_{AFE}$ HfO$_2$ structure determined from the present work.

| Space group | No. 61, *Pbca* | |
|---|---|---|
| | Relaxed structure | HRTEM measurement |
| *a* | 10.11153 Å | 10.10 Å |
| *b* | 5.13439 Å | 5.17 Å [a)] |
| *c* | 5.28054 Å | 5.29 Å |
| alpha | 90º | |
| beta | 90º | |
| gamma | 90º | |
| Volume | 274.147 Å³ | 276 Å³ |
| Atomic fractional coordinate | | Wyckoff |
| Hf | 0.1377, 0.0414, 0.6557 | 8*c* |
| O1 | 0.0317, 0.3242, 0.8393 | 8*c* |
| O2 | 0.2243, 0.7476, 0.4111 | 8*c* |
| Total energy[b)] | 26 meV f.u.$^{-1}$ | |

[a)]The ***b*** axis is not visible in the electron beam direction. Hence, the experimental value of the *b* lattice parameter of the *M* structure from ICSD 27313 was assumed. Owing to the coherent intergrowth of the $O_{FE}$ and *M* phases, the value of the *b* lattice parameter of the $O_{AFE}$ structure is the same as for the *M* structure. This value was used to estimate the unit cell volume; [b)]Relative to the monoclinic *P*2$_1$/*c* (*M*) phase.

**Table S6. Lattice parameters for the *M* phase $HfO_2$.**

| Space group | No. 14, $P2_1/c$ | | |
|---|---|---|---|
| | Relaxed structure | ICSD 27313 PDF 34-104 | HRTEM measurement |
| *a* | 5.10409 Å | 5.1126 Å | 5.11 Å |
| *b* | 5.14964 Å | 5.1722 Å | 5.17* Å |
| *c* | 5.29164 Å | 5.2948 Å | 5.29 Å |
| alpha | 90° | 90° | |
| beta | 99.64° | 99.18° | 99.56° |
| gamma | 90° | 90° | |
| Volume | 137.123 Å$^3$ | 138.22 Å$^3$ | 138 Å$^{3\,a)}$ |
| Atomic fractional coordinate | | Wyckoff | |
| Hf | 0.279869, 0.041855, 0.217835 | 4*e* | |
| O1 | 0.066657, 0.321647, 0.366317 | 4*e* | |
| O2 | 0.456028, 0.744544, 0.499698 | 4*e* | |

a)The *b* axis is not visible in the electron beam direction. Hence, the experimental value of the *b* lattice parameter of the *M* structure from ICSD 27313 was assumed and used to estimate the unit-cell volume.

**Table S7. Lattice parameters for the $O_I$ phase HfO$_2$.**

| Space group | No. 61, *Pbca* | |
|---|---|---|
| | Relaxed structure | Experiment |
| *a* | 9.98228 Å | 10.0177 Å |
| *b* | 5.20834 Å | 5.2276 Å |
| *c* | 5.04843 Å | 5.0599 Å |
| *alpha* | 90º | |
| *beta* | 90º | |
| *gamma* | 90º | |
| *Volume* | 262.474 Å$^3$ | 264.980 Å$^3$ |
| Atomic fractional coordinate | | Wyckoff |
| Hf | 0.3843, 0.0358, 0.2458 | 8*c* |
| O1 | 0.2887, 0.3720, 0.3755 | 8*c* |
| O2 | 0.0225, 0.2387, 0.0020 | 8*c* |
| Total energy[a] | 80 meV f.u.$^{-1}$ | |

[a] Relative to the monoclinic $P2_1/c$ (*M*) phase.

**Table S8. Relaxed lattice parameters for the $O_{AFE}$ phase of ZrO$_2$.**

| | |
|---|---|
| Space group | No. 61, *Pbca* |
| *a* | 10.28093 Å |
| *b* | 5.22831 Å |
| *c* | 5.36541 Å |
| *alpha* | 90º |
| *beta* | 90º |
| *gamma* | 90º |
| Volume | 288.401 Å$^3$ |

| Atomic fractional coordinate | | Wyckoff |
|---|---|---|
| Hf | 0.1377, 0.0414, 0.6557 | 8*c* |
| O1 | 0.0317, 0.3242, 0.8393 | 8*c* |
| O2 | 0.2243, 0.7476, 0.4111 | 8*c* |
| Total energy[a] | 27 meV f.u.$^{-1}$ | |

[a]Relative to the monoclinic $P2_1/c$ (*M*) phase.